\documentclass[aps,prd,twocolumn,nofootinbib,superscriptaddress,showpacs,amsmath]{revtex4-1}
\usepackage{graphicx,epsf,amssymb,url,hyperref}
\usepackage{mathrsfs}
\usepackage[]{latexsym}
\usepackage[UKenglish]{babel}
\usepackage{epsfig}
\usepackage[utf8]{inputenc}

\newcommand{\be}{\begin{equation}}
\newcommand{\ee}{\end{equation}}

\def\comment#1{}

\begin{document}

%%%%%%%%%%%%%%%%%%%%%%%%%%%%%%%%%%%%%%%%%%%%%%%%%%%%%%%%%%%%%%
\title{$ER= EPR$ and Non-perturbative action integrals for quantum gravity}
%%%%%%%%%%%%%%%%%%%%%%%%%%%%%%%%%%%%%%%%%%%%%%%%%%%%%%%%%%%%%%
%1st author
\author{Salwa~Alsaleh}
\email{salwams@ksu.edu.sa}
\affiliation{ Department of Physics and Astronomy, King Saud University, Riyadh 11451, Saudi Arabia}
%
%2nd author 
\author{Lina Alasfar}
\email{lina.alasfar@outlook.fr}
\affiliation{ Universit\'{e} de Clermont Auvergne, 24 Avenue des Landais F-63177 Aubi\`{e}re Cedex, France }
%
%%%%%%%%%%%%%%%%%%%%%%%%%%%%%%%%%%%%%%%%%%%%%%%%%%%%%%%

\date{\today}
            
%%%%%%%%%%%%%%%%%%%%%%%%%%%%%%%%%%%%%%%%%%%%%%%%%%%%%%%%%%%%%%%%%%%%%%%%%%%%
\begin{abstract}
In this paper, we construct and calculate a non-perturbative path integrals in a multiply-connected space-time. This is done by by summing over homotopy classes of paths. The topology of the space-time is defined by Einstein-Rosen bridges (ERB) forming from the entanglement of   quantum foam described by virtual black holes.  
As these `bubbles' are entangled, they are connected by Plankian ERB's because of the  $ER=EPR$ conjecture.  Hence the space-time will possess a large first Betti number~$B1$. For any compact 2-surface in the space-time, the
topology (in particular the homotopy) of that surface is not-trivial, due to the large number of Plankian ERB's that define homotopy though this surface. The quantisation of space-time with this topology - along with the proper choice of the 2-surfaces - is conjectured to allow anon-perturbative path integrals of quantum gravity theory over the space-time manifold.
\end{abstract}
%%%%%%%%%%%%%%%%%%%%%%%%%%%%%%%%%%%%%%%%%%%%%%%%%%%%%%%%%%%%%%%%%%%%%%%%%%%%%%%%
\maketitle
%%
%\tableofcontents
%%
%%%%%%%%%%%%%%%%%%%%%%%%%%
\section{Introduction}
%%%%%%%%%%%%%%%%%%%%%%%%%%
%
%
\par \noindent In the past few years, a new approach in understanding and constructing quantum geometry had emerged from linking geometric quantities to information-theoretic one. This idea first emerged from the calculation of entropy of spherically symmetric isolated horizons~(SSIH) in Loop quantum gravity~\cite{immirzi1997real,PhysRevLett.80.904}.  Later, the emergence of holographic principle to include the `entanglement entropy' of the vacuum to be linked to the are of the surface enclosing a geodesic-generated ball~\cite{reznik2003entanglement}. A deeper link link between quantum geometry and information theory was established~\cite{ER1}, as a solution to AMPS / Firewall paradox~\cite{almheiri2013black}. This solution suggested that for two black holes collapsed from an entangled matter, these black holes would be connected by an Einstein-Rosen Bridge~(ERB). This conjecture became widely known as the~$ER= EPR$ \textit{conjecture}. One can go even further with conjecture and claim that for any pair of entangled particles, there is a Plankian ERB, that matter will not fall into it. However, this can considered in an abstract geometric sense~\cite{ER2}.
 \par \noindent  The development of this conjecture has raised the possibility that space-time could be interwoven by entanglement, and wormholes could be holding the space-time together,  This idea has been applied in category theory and topological quantum field theory to show that quantum fluctuations correspond to formation of wormholes and vice versa~\cite{baez2014wormholes}. Technically speaking, the~$ER=EPR$ raises  the possibility that space-time could multiply connected. However, how one could start doing physics in multiply-connected space-time ? More importantly, how to formulate the physic of multiply connected space-time; and would it help to do so ?
 \par \noindent The most basic way to start is by replacing the points of the space-time manifold~$ \mathcal{M}$ by a non-Abelian algebra generators,establishing non-commutative geometry~\cite{Nicolini:2005vd,Gingrich:2010ed},
\begin{equation}
\label{qgeo}
[q^\mu , q^\nu] = i \ell_p^2 g^{\mu \nu}.
\end{equation}
This is a direct result of combining gravitation theory with uncertainty relation of quantum mechanics. It is immediately recognised that~$ \ell_p$ is the Plank length, and the relation~\eqref{qgeo} implies that space-time at the Plank scale consists of what-so-called quantum foam~\cite{wheeler1}. First described by J. Wheeler and later studied in detail by S. Hawking~\cite{hawking1996virtual}, this quantum foam consists of a sea of virtual black holes, which Hawking described them as~\textit{bubbles}. These bubbles make the space-time posses a large second Betti number. Moreover, in his paper, Hawking argued that the space-time is simply connected and admits a topology~$ S^2 \times S^2$ or~$ K_3 \times K_3$ \dots etc. 
\par \noindent Nevertheless, these bubbles can form the multiply-connected space-time we are looking for if we considered the~$ ER= EPR $ conjecture. In this paper, we shall discuss how to define a multiply-connected space-time combing both ideas discussed above. Next, show how this picture could help in the theory of quantum gravity, after that an application for this conjecture is discussed at the hand-waving level by recovering the Bekenstein-Hawking formula for Black hole entropy. Finally, we will point out the possible steps for this programme that ought to be taken in order to make the argument in this paper more formal.
%%
%%%%%%%%%%%%%%%%%%%%%%%%%%%%
\section{The space-time at the Plank scale}
%%%%%%%%%%%%%%%%%%%%%%%%%%%%
%%
\par \noindent Let~$\mathcal{M}$ be a 4-dimensional~$C^k$-manifold that admits a  topology~$ \mathbb{R} \times \Sigma$, we deal with globally hyperbolic space-time in order to prepare for canonical formalism of gravity. If any field was put on~$ \mathcal{M}$,  by assigning smooth sections~$ \varsigma(t,x) ; t,x \in \mathbb{R} \times \Sigma$ over~$\mathcal{M}$ to an associated bundle~$ \mathcal{F}$. This field is quantised by standard canonical quantisation rules, taking the field to be a scalar field~$ \phi (t,x) = \varsigma(t,x)$, the quantisation for the field is given by the commutation relation for~$\phi$ and its conjugate momentum~$ \pi (t,x)$ which is itself a section ~$ \varpi(t,x)= \frac{d\varsigma(t,x)}{dt} $,
\begin{align}
[\phi(t_0, x), \phi(t_0,x')] = 0, \nonumber \\
[\pi(t_0, x), \pi(t_0,x')]= 0 ,\nonumber \\
[\phi(t_0, x), \pi(t_0,x')]= i \delta(x-x').
\end{align}
These rules imply a Fock space \footnote{ Since we are dealing with quantum field theory on curved manifold, the Fock space and vacuum spates are by no means unique} and thereby we can define a Hamiltonian~$H$, whose expectation value over any vacuum state diverges as it is known from standard quantum field theory. The only non- perturbative method to remove such divergence is to impose some UV-cutoff on the modes of the filed. This cutoff condition can be made because of the gravity theory present on~$\mathcal{M}$. For general relativity~(GR) one can find this cutoff term to be for modes with wavelengths less than or equal to the plank length $ \ell_p =1$ in Plankian units
In the semi-classical GR, the geometry itself fluctuates like the quantum fields that lie on it to show this take a small 3-space like surface $\sigma$, let it be a geodesic-generated ball with radius $l$. We now integrate Einstein field equations over this ball~(in some chart defined over that small ball),
\begin{equation}
\frac{1}{4\pi} \int_\sigma d\sigma^\nu G_{\mu\nu} = 2 \int_\sigma d\sigma^\nu \langle\ \hat{T}_{\mu\nu}\rangle .
\end{equation} 
The LHS is equal to the radius of curvature domain of small region of space time~$ R_\mu $. For the RHS, we consider the stress energy tensor of the scalar field~$ \phi$ we get,
\begin{equation}  
\label{radofcurve}
R_\mu =\int d^3k \left( \omega_k +k_i\right) \langle 0| \hat{a}^\dagger_k \hat{a}_k+ \hat{a}_k\hat{a}^\dagger_k |0\rangle.
\end{equation}
The operators~$ \hat{a}^\dagger_k ,\hat{a}_k$ are the creation and inhalation operators of the defined Fock space. Here, the Latin indices run from $1$ to $3$ and the Greek ones run from $0$ to $3$, and $ \omega_k$ is the normal mode given by $\sqrt{k_j k^j}$. The expression~\eqref{radofcurve} can be further treated yielding,
\begin{equation}  
\label{div}
R_\mu =\int d^3k \sum_{k=0}^{\infty} \omega_k
\end{equation}
Thus the radius of curvature domain is divergent, implying that the geometry is not supposed to exist with field fluctuating like that. Hence, locality and flat geometry cannot coexist. To get rid of that divergence we use the uncertainty relation~$ \Delta R_\mu x_\mu \sim 1$ . This is derived from Hawking's assumption about the quantum structure of space-time at the micro scale~\cite{hawking1996virtual},  as a sea of virtual blackholes. By this assumption, we cannot define the field nor the geometry at one point, but rather a region with minimal length. Thus geometry fluctuation is a key element in the stability of classical geometry. Concluding geometry must me quantised, i.e. the space-time is made from discrete bubbles of geometry. This result is obtained without  assumptions outside GR and quantum theory.Virtual black holes were also studied in third quantised formalism of canonical quantum gravity~\cite{faizal2012some}, in dilatorn gravity theories~\ref{Grumiller:2002dm,Grumiller:2001rg,Grumiller:2000ah}, and in other alternative theories as in generalised uncertainty principle and their phenomenological implications in proton decay~\cite{alsaleh2017virtual} and higher spin  theories from Teukolsky equations \cite{Prestidge:1998bk}. Virtual black holes are also important in understanding the black hole information paradox~\cite{Calmet:2014uaa}, for example, it has been shown that the S-matrix of scattering processes by virtual black hole is incoherent~\cite{Hawking:1997ia}, leading to a possible indication that information paradox could be resolved by studying the phenomenology of scattering by virtual black holes~\cite{Grumiller:2004yq}.
\par \noindent The predictions of quantum mechanics, these nearby virtual black holes are entangled. In order to show this, consider  black holes in a box , each black hole will emit Hawking radiation to be absorbed by the others, even when the system reaches thermal equilibrium, this process will continue to happen. If we assign a density matrix for this system, it will evolve to a mixed states density matrix over time, even if we started with a pure state (each state describes a Black hole.). Viz., black holes will get entangled by absorbing each other's Hawking radiation~\cite{ER1}. 
\par \noindent Similar argument can be made to the virtual black holes, that make up the quantum foam. Another way to show that the quantum foam is indeed entangled, and hence is filled with Plankian Wormholes, is the work done by~\cite{baez2014wormholes}, using category theory, they concluded that the pair production of quantum fluctuation correspond to formation of a wormhole. Although their argument was for topological field theories, it can be proven that this argument could be extended to general relativity if we considered it as a form of BF theory. Identifying the form $B =\ast [e\wedge e]$, where $e$ are the triebein fields corresponding to the 3 metric $\gamma$ of $\Sigma$, then the Palatini action reads,
\begin{equation}
S_P = \int_{\Sigma} \text{Tr} (\ast [e\wedge e] \wedge F ).
\end{equation}
However, it remains to be proven explicitly that these results will  hold for full relativistic gravity theory.
\par \noindent Assigning to each unit volume on the 3 spacelike surface~$ \Sigma$ a quantum state~$ | \psi\rangle$, this unit area can carry only 1 bit of information, either it contains a virtual black hole~$| 1\rangle$ or not~$|0 \rangle$ Hence, we can write the quantum state for such unit area,
\begin{equation}
| \psi\rangle = \frac{1}{\sqrt{2}} \left(  | 1\rangle +  |0\rangle\right).
\end{equation}
We observe that given any 2-surface~$ \Delta$, enclosing a region of~$ \Sigma$ , we can establish near the boundary states of the volumes in the `in' and `out' regions. These states describe entangled states,
\begin{equation}
| \Psi\rangle = \frac{1}{\sqrt{2}} \left(  | 1_{in}\rangle \otimes |1_{out}\rangle+ | 0_{in}\rangle \otimes |0_{out}\rangle\right).
\end{equation}
The state~$ | \Psi\rangle$ belongs to the Hilbert space~$ \mathcal{H} = \mathcal{H}_{in} \otimes \mathcal{H}_{out}$ of the space-time volumes.
This construction of space-time from entanglement does not require holography as in~\cite{van2010building}, but basically shares the same aim. In order to do so, it is needed to define the topology of the space-time having this construction.
%%
%%%%%%%%%%%%%%%%%%%%%%%%%%%%
\section{ The topology of the entangled quantum foam}
%%%%%%%%%%%%%%%%%%%%%%%%%%%%
%%
\par \noindent Since these volume elements~(making up) the quantum foam are entangled, using the $ER=EPR$ conjecture, each bubble of geometry is connected to another by a wormhole/ ERB. Making the space-time multiply connected. However, in 3 manifolds, we cannot clearly define the first Betti number corresponding the the first fundamental group. The first fundamental group, appears (well defined) for 2 surfaces. Taking the surface~$\Delta$ in the previous section, and analysing its homotopy we find that, and due to  the $ER=EPR$. We can define loops between the `in' and `out' regions that cannot be shrunk to a point. These loops are made by passing through the wormholes made by the entangled bubbles. The surface~$\Delta$ is them homeomorphic to an~$N$-punctured sphere, $N+1$ being proportional to the area of the surface in Plank-area units. 
\par \noindent The first fundamental group for the surface~$\Delta$ is generated by $N$ generators, each go around each puncture made by a wormhole. Note that we should only consider bubbles entangled in the region~$ Ar (\Delta) \pm 1$, the entanglement for bubbles further from Plank length is negligible. Hence the rank of this free group is proportional to~$ Ar( \Delta)$. The calculation of the proportionality constant is rather difficult, because we need to take into an account that the classical area is obtained only by filling the surface with infinite circles, while in quantum geometry, the 'quantum' area is a bit less, due to packing factor~(see Tammes Problem, Ref~\cite{tammes1930origin}). The exact number of generator is the number of wormholes forming between bubbles filling the area~$ Ar( \Delta-1)$ and~$ Ar( \Delta+1)$, assuming all the black holes in these two layers are entangled, the number of the wormholes forming is the area of the surface~$\Delta$ that comes between the layers multiplied by a filling factor, i.e. how many bubbles one can fit in the volume covered by~$ Ar( \Delta+1)$ minus the volume covered by~$ Ar( \Delta-1)$. 
%%
%%%%%%%%%%%%%%%%%%%%%%%%%%%%
\section{Calculation of path integrals }
%%%%%%%%%%%%%%%%%%%%%%%%%%%%
%%
\par \noindent The topology of space-time could allow a non perturbative calculation of path integrals, automatically removing divergences that appear in the standard path integrals. Since we have found$ \mathcal{M}$ to be multiply connected, we can look at its universal covering space, and preform path integration on the covering space. Instead of integrating over all paths of the same homotopy class, one can instead integrate over homotopy classes in the universal cover.  Nevertheless, this is not helpful in unbounded spaces, as one still gets infinite classes to integrate over. Rather, one may take two points in~$\mathcal{M}$, $ p,q$. Surround $p$ with a 2-spacelike surface~$ \sigma_p$ and~$ q$ with another~$ \sigma_q$. Homotopy classes of paths between~$p$ and~$q$ will be finite due to the topology of the surfaces $\sigma_p$ and $ \sigma_q$. One can preform a non perturbative integration over the homotopy classes. 
\par \noindent Nevertheless, there is a rather peculiar feature of this technique, that is the dependence of the value of the path integral of the choice of surfaces. It remains an unsolved problem that needs special attention.  This `freedom' in choosing the surfaces, could correspond to gauge freedom, that  one can mod out. Alternatively, there is no freedom and it is fixed by the theory of gravity, like the case with the horizon of a black hole, this surface is fixed by gravity, and one needs to use - only- for the calculation of action on that surface. 
Hence the general technique for calculating path integrals in a multiply-connected space-time could be summarised in the following points,
%%%
%
\begin{itemize}
	\item Pick two points in the space-time manifold $\mathcal{M}$.
	\item  Surround these points by 2 spacelike surfaces.
	\item Identify the homotopy classes and loops on these surfaces, then left them up to the associated bundle in which the field of concern 'lives'. Such than one can preform parallel transport along the loops.
	\item Define the fundamental groups for the surfaces, and find a proper representation for these groups (loop representation).
	\item Using the above representation, preform the standard technique for path integration in multiply connected space-time,
	\begin{equation}
	\int \mathcal{D}[f] e^{ -i S[f, f,_\mu]}\longrightarrow \sum_{\alpha \in \pi_1 (\sigma_p \cup \sigma_q)} \chi(\alpha) K^\alpha.
	\end{equation}
	Where $ \alpha$ is the homotopy class, $ \chi$ is a representation of the fundamental group and $ K^{\alpha}$ is the transition amplitude via the homotopy class $ \alpha$. 
\end{itemize}
%%
%
%%
%%%%%%%%%%%%%%%%%%%%%%%%%%%%
\section{Conclusion}
%%%%%%%%%%%%%%%%%%%%%%%%%%%%
%%
\par \noindent The action path integrals over a multiply connected spaces could be shown to be non-peturbative and yield exact results as one might sum over equivalent classes of paths that have the same winding number. In fact, in this paper, we have seen that merging the ideas of minimal length and creation of pair black holes~\cite{hawking1996virtual}, along  with the ER=EPR conjecture \cite{ER1,ER2} could yield a multiply-connected space-time. Hence path integrals over space-time with gravity could be non-peturbative.
\par \noindent It could be shown in greater detail that the creation and inhalation of black-holes as a result of third-quantised theory of geometry~\cite{faizal2012some}. Hence one could also apply the non-peturbative techniques to geometry itself. There is a lot to add to this proposal in order to use it for exact calculation of path integrals  , and/or quantisation of geometry. Based on the microscopic structure of space-time, and to generalise the calculations in topological field theory~\cite{baez2014wormholes} to the 4D relativistic space-time, that is non-topological. 
\par \noindent  As the topology is scale-independent,the non trivial topology of scape-time caused by the virtual wormholes forming the quantum foam, could have a low-energy effects. It  would be interesting to investigate such effects in future work, such as the Aharonov-Bohm effect, corrections to scattering amplitudes in QED and QCD and shifts in energy levels of nuclear and atomic systems.
%%
%%%%%%%%%%%%%%%%%%%%%%%%%%%%%%%
%%%%%%%%%%%%%%%%%%%%%%%%%%%%%%%
\section*{Acknowledgements}
%%%%%%%%%%%%%%%%%%%%%%%%%%%%%%%
%%%%%%%%%%%%%%%%%%%%%%%%%%%%%%%
{ \fontfamily{times}\selectfont
	\noindent 
	This research project was supported by a grant from the `` Research Center of the Female Scientiffic and Medical Colleges'' , Deanship of Scientiffic Research, King Saud University. 
	\bibliography{bib}
	\bibliographystyle{iteer}
\end{document}